\documentclass[a4paper,11pt]{article}
\usepackage{pos}
\usepackage{caption}
\usepackage{subcaption}

\usepackage{comment}

\renewcommand{\logo}{\relax}

\title{Lensing effects of ultra-high-energy cosmic rays propagating in the Galactic magnetic field}
\ShortTitle{Lensing effects of UHECRs propagating in the Galactic magnetic field}

\author*[a]{Danelise de Oliveira Franco}
\author[a]{Günter Sigl}

\affiliation[a]{Universität Hamburg, II. Institut für Theoretische Physik,\\
  Luruper Chaussee 149, Hamburg, Germany}

\emailAdd{danelise.franco@desy.de}
\emailAdd{guenter.sigl@desy.de}

\abstract{Since ultra-high-energy cosmic rays (UHECRs) are electrically charged particles, they are deflected by magnetic fields. Those magnetic fields can act as lenses, altering their trajectories and (de)magnifying their apparent source images. These deflections of UHECR trajectories can lead to phenomena such as the appearance of multiple images of an extragalactic source. In this study, we investigate the influence of the Galactic Magnetic Field (GMF) on the propagation of UHECRs, considering four different realistic models for the GMF: the \texttt{PT11}, the \texttt{JF12}, the \texttt{UF23}, and the \texttt{KST24} models. We investigate how an isotropic flux on Earth would have entered the edge of the Galaxy for different rigidity values from 1 to 100\,EV ($\equiv 10^{18}$\,V). In addition, we investigate the appearance of multiple images of astrophysical point sources. Furthermore, we analyze the modification of the cosmic ray flux from a source as a function of the rigidity and its dependence on the chosen GMF model. Since the deflection induced by the magnetic field depends on the rigidity of the particle, the effects vary among different nuclear species. Consequently, our findings can have implications for interpreting mass-composition and anisotropy observations, as the rigidity-dependent deflections directly alter the observed UHECR arrival direction distribution.}

\FullConference{39th International Cosmic Ray Conference (ICRC2025)\\
 15–24 July 2025\\
Geneva, Switzerland\\}

\begin{document}
\maketitle

\section{Introduction}

Ultra-high-energy cosmic rays (UHECRs) are conventionally understood as ionized nuclei with energies $E\gtrsim 1\,\mathrm{EeV}\equiv 10^{18}$\,eV. Deflections caused by the extragalactic and Galactic magnetic fields (EGMF and GMF) pose a challenge in identifying their sources. While the EGMF in the voids is estimated to be $\lesssim 1$\,nG \cite{batista2021intergalactic}, magnetic fields within spiral galaxies similar to the Milky Way can reach a few $\mu$G and extend over several kiloparsecs~\cite{beck2015spiral}. Since magnetic fields affect the trajectories of the UHECRs, a robust understanding of the GMF structure is essential for charged-particle astronomy. 

In recent years, significant progress has been made in the UHECR field, thanks to data recorded by the two largest cosmic-ray observatories in the world: the Pierre Auger Observatory~\cite{auger2015observatory} and the Telescope Array experiment~\cite{abuzayyad2012telescopearray}. A milestone was the observation of a dipolar anisotropy signal by the Pierre Auger Collaboration, pointing $125^{\circ}$ away from the Galactic center~\cite{aab2017anisotropy}, indicating a possible extragalactic origin of the highest energy cosmic rays. However, accurately interpreting large-scale anisotropies requires careful consideration of the deflections caused by both the EGMF and the GMF~\cite{bakalova2023dipole, bister2024anisotropy, hackstein2018anisotropy, rossoni2025anisotropy}.

Since the deflection caused by magnetic fields depends on the rigidity of the particle, $R = E /Z$ (where $E$ is the energy of the particle and $Z$ is the charge number), the EGMF and the GMF play an important role in the propagation of UHECRs. Consequently, magnetic fields affect different species with the same energy distinctly, leading to a dependence on mass composition and impacting the arrival direction distribution (see e.g.,~\cite{mollerach2025magnetic}). Recent inferences of the depth of the shower maximum, $X_{\mathrm{max}}$,  indicate that the mass composition becomes heavier with increasing energy~\cite{Abdul2025masscomposition}. For the past few years, the Pierre Auger Observatory has been undergoing an upgrade named \textit{AugerPrime}~\cite{castellina2019prime}. One of the primary goals of this upgrade is to enhance the mass composition sensitivity, gathering new information on an event-by-event basis, thereby potentially shedding light on the mass-dependent deflections caused by cosmic magnetic fields.

The coherent component of the GMF is primarily probed through two astrophysical observations: the Faraday rotation measures of extragalactic polarized radio sources and the polarized intensity of the synchrotron emission of cosmic-ray electrons and positrons within the Galaxy. Those two quantities provide complementary information about the GMF, as the former provides information about the parallel and the latter about the perpendicular components to the line of sight of an observer on Earth. In the last two decades, new measurements of both quantities have significantly advanced the development of more realistic GMF models.

In this work, we investigate UHECR propagation through four different GMF models, all implemented within the \texttt{CRPropa 3.2} framework~\cite{batista2022crpropa}. We first consider the model developed by Pshirkov, Tinyakov, Kronberg, and Newton-McGee and published in 2011 (hereafter \texttt{PT11})~\cite{pshirkov2011magnetic}, which employs a bisymmetric disk field with the halo component adopted from Sun et al.~\cite{sun2008radio}. We also consider the Jansson and Farrar model published in 2012 (hereafter \texttt{JF12})~\cite{jansson2012magnetic}, one of the most widely used GMF models for UHECR propagation studies. For this work, we investigate only its large-scale regular field component, although the \texttt{JF12} model also allows the inclusion of striated and small-scale random fields. Furthermore, we extend our analysis to two recently released models: the Unger and Farrar model (hereafter \texttt{UF23})~\cite{unger2024magnetic} and the Korochkin, Semikoz, and Tinyakov model (hereafter \texttt{KST24})~\cite{korochkin2024magnetic}. The \texttt{UF23} model has eight variants, and for our study, we utilize the fiducial variant, called \texttt{base}, which is the most data-driven.

The GMF can act as a giant cosmic lens by deflecting the trajectories of UHECRs as they propagate through the Galaxy until reaching Earth, therefore possibly (de)focusing their apparent source images. Early studies in the 2000s extensively investigated these lensing effects~\cite{harari1999spectrum, harari2002lensing}. We examine the impact of the GMF models considered in this work, analyzing how they deflect particles, potentially generate multiple images of a single source, and (de)magnify apparent source regions on the sky. Given the rigidity-dependent nature of these deflections, we explore these effects across a range of UHECR rigidities, thus providing insight into how different mass groups might be affected.

\section{UHECR deflections}

The deflection angle of a UHECR is defined as the angle between its direction when entering the Galactic halo and its observed arrival direction on Earth, after propagation through the GMF. Performing a numeric forward propagation simulation of particles from the Galactic halo to Earth presents a significant computational challenge, as most of the injected UHECRs do not intercept Earth, making the process inefficient for collecting enough statistics. To avoid this issue, a widely adopted technique is to backtrack particles. The backtracking is performed by propagating the antiparticle, injecting it on Earth, until it reaches the Galactic halo. This is equivalent to forward tracking the particle from the halo to Earth, but ensuring that all simulated particles arrive on Earth. 

For this work, the edge of the Galaxy is modeled as a sphere, centered at the Galactic center, with a radius of 20\,kpc, a distance at which GMF effects are typically negligible.\footnote{For the \texttt{UF23} model, we consider a 30\,kpc radius sphere.} The simulations are executed using framework \texttt{CRPropa 3.2}~\cite{batista2022crpropa}. To define a uniform arrival direction distribution on Earth, we pixelize the sky using Healpy~\cite{zonca2019healpy}, a Python library based on HealPix~\cite{gorski2005healpix}, using $N_{\mathrm{side}} = 128$ and take the center of each pixel as the injected momentum direction, resulting in 196,608 directions.\footnote{The number of pixels is defined by the parameter $N_{\mathrm{side}}$ by the formula $N_{\mathrm{pixels}} = 12 \cdot N_{\mathrm{side}}^{2}$.}

\begin{figure}[ht!]
    \centering
    \includegraphics[width=\textwidth]{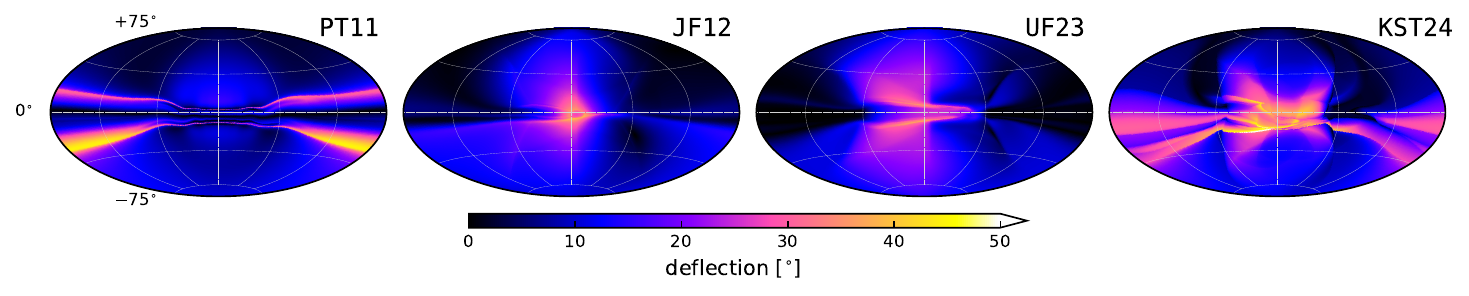}
    \caption{Angular deflections for different GMF models at a rigidity of $R = 30$\,EV as a function of the arrival direction at Earth. The skymaps are presented in Galactic coordinates using the Hammer projection.}
    \label{fig:deflections}
\end{figure}

Several works studied the deflections caused by the GMF and their dependence on the chosen model (see e.g.,~\cite{farrar2019deflections, korochkin2025deflections, unger2025deflections}). We illustrate the differences in Fig.~\ref{fig:deflections}, where we show the deflection angles obtained for the four models considered in this work for the rigidity $R = 30$\,EV in Galactic coordinates. In all skymaps throughout this work, the Galactic longitude increases to the left. As shown, different GMF models lead to different deflection magnitudes. This consequently implies variations in the effects on the source images, as we will explore in the next sections.

\section{Multiple images of a source}

Magnetic deflections can significantly alter the trajectories of the UHECRs, potentially causing the appearance of multiple images from a single extragalactic source. To study this effect, we model an isotropic UHECR arrival direction distribution at Earth and investigate how this distribution would have entered the Galactic halo.  We produce an isotropic arrival direction distribution by pixelizing the sky using $N_{\mathrm{side}} = 256$ and taking the center of each pixel to define the injected momentum direction, resulting in a total of 786,432 directions. Then, we backtrack these particles through the GMF to determine their arrival directions at the edge of the Galaxy. To quantify the redistribution of particles, we define the variable $\mu(\boldsymbol{\hat{n}}_{i})$ as the ratio between the number of particles within a given pixel in the halo and the number of particles in the corresponding pixel on Earth,

\begin{equation}
    \mu(\boldsymbol{\hat{n}}_{i}) = \frac{N_{\mathrm{halo}} (\boldsymbol{\hat{n}}_{i})}{N_{\mathrm{Earth}} (\boldsymbol{\hat{n}}_{i})}.
\end{equation}

\begin{figure}[ht!]
    \centering
    \includegraphics[width=\textwidth]{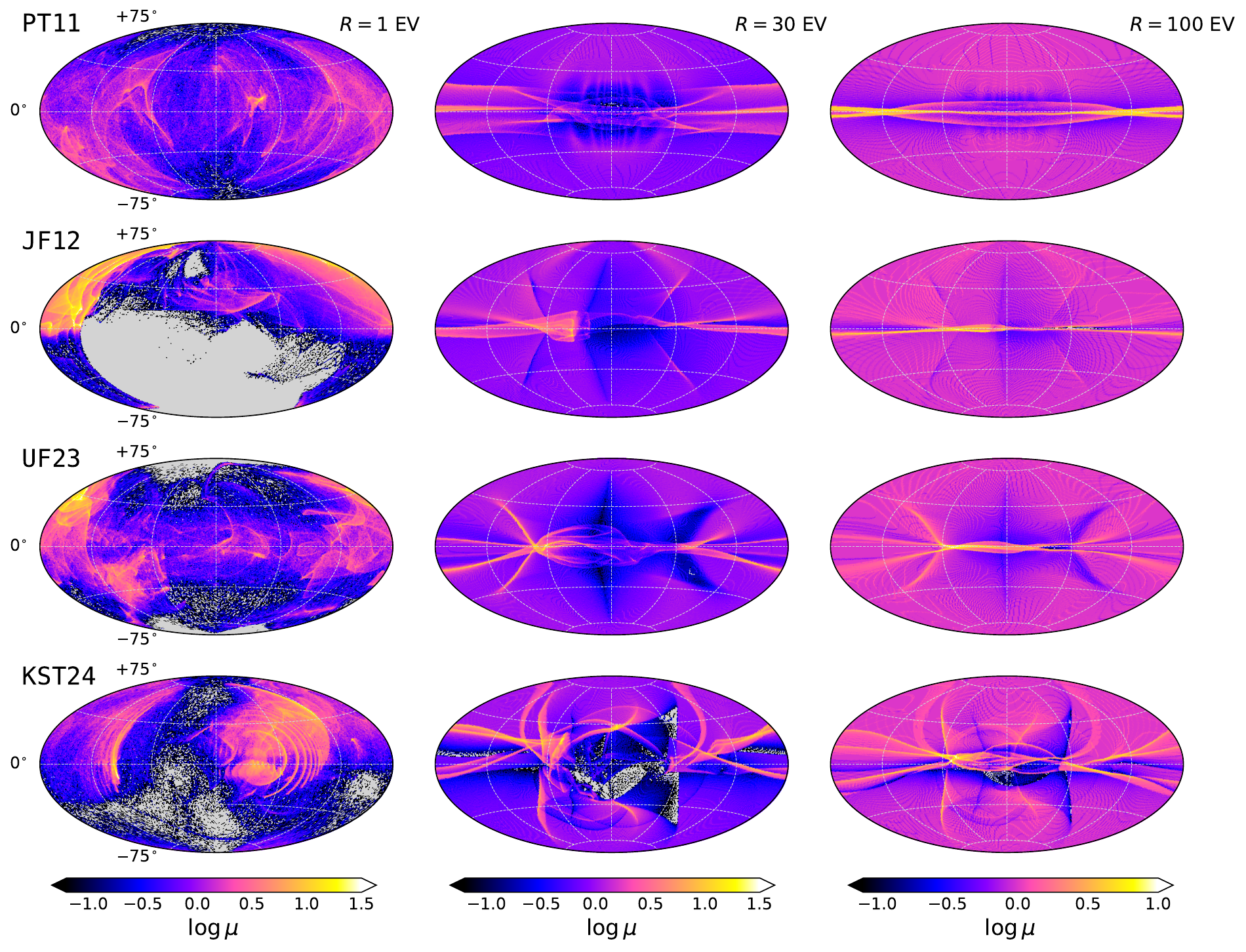}
    \caption{The ratio, $\mu(\boldsymbol{\hat{n}}_{i}) = N_{\mathrm{halo}} (\boldsymbol{\hat{n}}_{i}) /N_{\mathrm{Earth}} (\boldsymbol{\hat{n}}_{i})$, between the number of arrival directions on the Galactic halo and on Earth as a function of the arrival direction at Earth. Values are presented on a logarithmic scale in Galactic coordinates. We mask pixels with no particles in the halo, showing them in gray. Each row represents a different GMF model, and each column a distinct rigidity value. Maps generated using 786,432 simulated particles and $N_{\mathrm{side}} = 64$ (16 particles per pixel for an isotropic distribution).}
    \label{fig:sheets}
\end{figure}

Figure~\ref{fig:sheets} presents the results for particles propagating through various GMF models at three different rigidity values: $R=1$, 30, and 100\,EV, as a function of the arrival direction at Earth. The plots are produced using $N_{\mathrm{side}} = 64$, which means that the arrival direction distribution on Earth contains 16 particles per pixel. Farrar and Sutherland used a similar approach to build magnification maps~\cite{farrar2019deflections}. As we can see in Fig.~\ref{fig:sheets}, these deflections can result in regions of high particle convergence, forming ``folded'' areas. These regions are indicative of the occurrence of multiple images of extragalactic sources. As expected, this effect is more intense at lower rigidity values. 

\begin{figure}[ht!]
     \centering
     \begin{subfigure}[c]{0.42\textwidth}
         \centering
         \includegraphics[width=\textwidth]{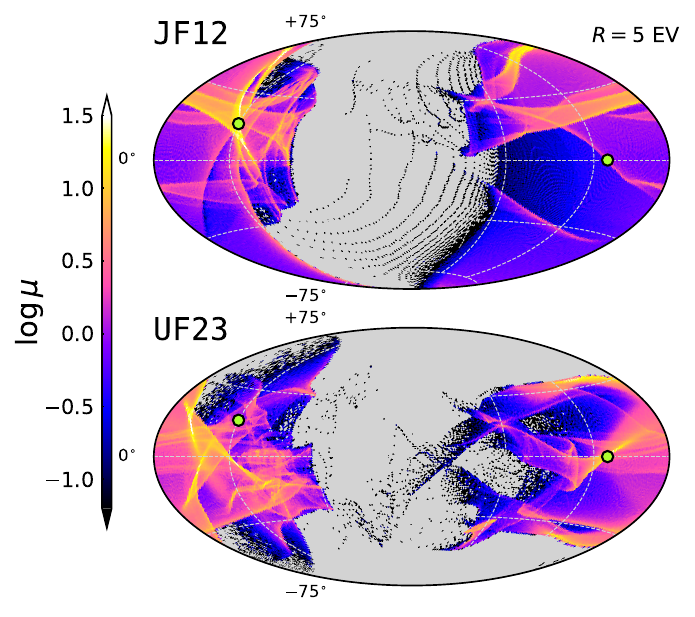}
     \end{subfigure}
     \begin{subfigure}[c]{0.57\textwidth}
         \centering
         \includegraphics[width=\textwidth]{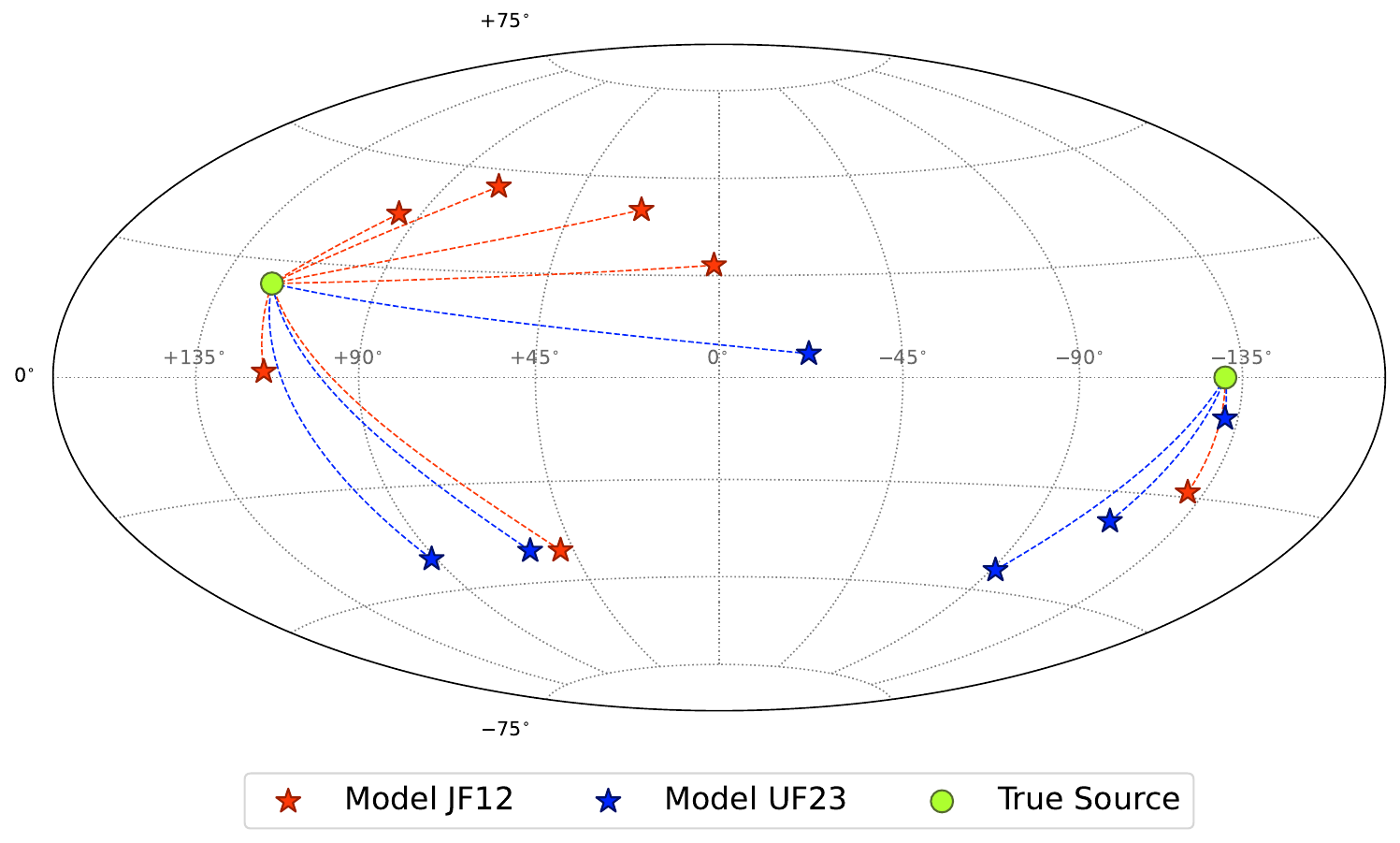}
     \end{subfigure}
\caption{On the left, the same as Fig. \ref{fig:sheets} for the rigidity 5\,EV and two different models (\texttt{JF12} and \texttt{UF23}). On the right, for the same rigidity, the multiple source images formed when UHECRs are propagated through these models, considering they come from two distinct fictitious sources located at ($\ell$, $b$) = ($120^{\circ}$, $20^{\circ}$) and ($-130^{\circ}$, $0^{\circ}$). For the first source, we have six different images when considering the \texttt{JF12} model and three for the \texttt{UF23} model, when for the second source we have three multiple images when considering the \texttt{UF23} model and a single image for the \texttt{JF12} model. The dashed lines connect the images (stars) to their respective source (circle).} 
\label{fig:multiple_images}
\end{figure}

Fig.~\ref{fig:multiple_images} illustrates in detail the phenomenon of multiple images for a single extragalactic source, considering a rigidity $R = 5$\,EV and considering two fictitious sources located at Galactic coordinates ($\ell$, $b$) = ($120^{\circ}$, $20^{\circ}$) and ($-130^{\circ}$, $0^{\circ}$).\footnote{In this work, $\ell$ and $b$ are the Galactic longitude and latitude, respectively.} To identify these observed images on Earth, we first determine the Earth-based pixels containing particles that, when backtracked, arrive in the Galactic halo within the pixel corresponding to the position of the source. To precisely locate the images on Earth, a finer search is performed around the centers of these identified pixels. This is done by backtracking additional particles with arrival directions around the pixel centers, allowing us to determine the exact source images (within the required accuracy) corresponding to the actual source. We performed this analysis using two distinct GMF models: the \texttt{JF12} and the \texttt{UF23}. The source located at ($120^{\circ}$, $20^{\circ}$) produces six different images when propagated through the \texttt{JF12} model, while three distinct images are produced when considering the \texttt{UF23} model. For the source located at ($-130^{\circ}$, $0^{\circ}$), we observe three different images with the \texttt{UF23} model, and a single image for the \texttt{JF12} model.

\section{CR flux modification from a point source}

The appearance of multiple images, discussed in the previous section, is intrinsically related to (de)magnification effects caused by the GMF. This can lead to an amplification or an attenuation of the CR flux from a source.  To investigate this effect, we follow the procedure described in the work of Harari, Mollerach, and Roulet~\cite{harari1999spectrum}. The method involves tracking the evolution of a small area, defined by a fiducial particle and two other particles initially displaced from it by small orthogonal vectors $\Delta \boldsymbol{x}_{k}$ ($k = 1, 2$), entering the Galaxy with parallel trajectories, as it propagates from the Galactic halo to Earth. This area in the halo (and on Earth) is defined as $A_{\mathrm{halo/Earth}} = (\Delta \boldsymbol{x}_{1} \times \Delta \boldsymbol{x}_{2}) \cdot \boldsymbol{\hat{p}}|_{\mathrm{halo/Earth}}$, where $\boldsymbol{\hat{p}}$ is the normalized momentum of the fiducial particle. 

The first step involves determining the position and momentum at which the fiducial particle enters the Galactic halo. To guarantee that this particle will eventually reach Earth, we backtrack its antiparticle, with charge $-Ze$ and energy $E$, by numerically solving the equation of motion,

\begin{equation}
    \frac{d^2 \boldsymbol{x}}{dt^{2}} = -\frac{Zec}{E} \frac{d\boldsymbol{x}}{dt} \times \boldsymbol{B}(\boldsymbol{x}),
    \label{eq:motion}
\end{equation}

\noindent
from Earth to the halo. We extract information on the magnetic field, $\boldsymbol{B}(\boldsymbol{x})$, using the fields implemented in the framework \texttt{CRPropa 3.2}. Upon reaching the halo, we reverse the momentum of the fiducial particle to initiate its forward tracking, along with the two displaced particles. However, rather than individually tracking all three particles, we follow the evolution of their displacement vectors by numerically solving the equation

\begin{equation}
    \frac{d^{2}\Delta \boldsymbol{x}_{k}}{dt^{2}} = \frac{Zec}{E} \left[\frac{d\boldsymbol{x}_{0}}{dt} \times \Delta \boldsymbol{B}_{k}(\boldsymbol{x}_{0}) + \frac{d \Delta \boldsymbol{x}_{k}}{dt} \times  \boldsymbol{B}(\boldsymbol{x}_{0}) \right],
\label{eq:displacement}
\end{equation}

\noindent
where $\Delta \boldsymbol{B}_{k} \equiv \boldsymbol{B}(\boldsymbol{x}_{0} + \Delta \boldsymbol{x}_{k}) - \boldsymbol{B}(\boldsymbol{x}_{0})$ and $\boldsymbol{x}_{0}$ is the trajectory of the fiducial particle. In the derivation of Eq.~\eqref{eq:displacement}, the term proportional to $\Delta \boldsymbol{x}_{k} \times \Delta \boldsymbol{B}_{k}$ is neglected, as it is of second order in small displacements. We follow the displacement until the point that the fiducial particle returns to Earth. We quantify the modification of the cosmic ray flux from a point source as the ratio between the area defined by the two displacement vectors on Earth and in the halo, $m(\hat{\boldsymbol{n}}_{\mathrm{image}}) =  A_{\mathrm{Earth}} (\hat{\boldsymbol{n}}_{\mathrm{image}}) / A_{\mathrm{halo}}(\hat{\boldsymbol{n}}_{\mathrm{source}})$, where $\hat{\boldsymbol{n}}_{\mathrm{image/source}}$ is the direction of the image/source.\footnote{In the work of Harari, Mollerach, and Roulet~\cite{harari1999spectrum}, they define the ratio $A_{\mathrm{halo}}/ A_{\mathrm{Earth}}$ as the \textit{magnification} of the CR flux from a point source.} In the work of Giacinti et al., they present a discussion of (de)magnification effects of a source~\cite{giacinti2010galactic}. In that work, they define an \textit{amplification factor}. This factor can be obtained by defining a small triangular area on Earth, using the arrival directions of three nearby particles, and backtracking this area to the halo. Then, the amplification is the ratio between the area on Earth and in the halo.

\begin{figure}[ht!]
    \centering
    \includegraphics[width=\textwidth]{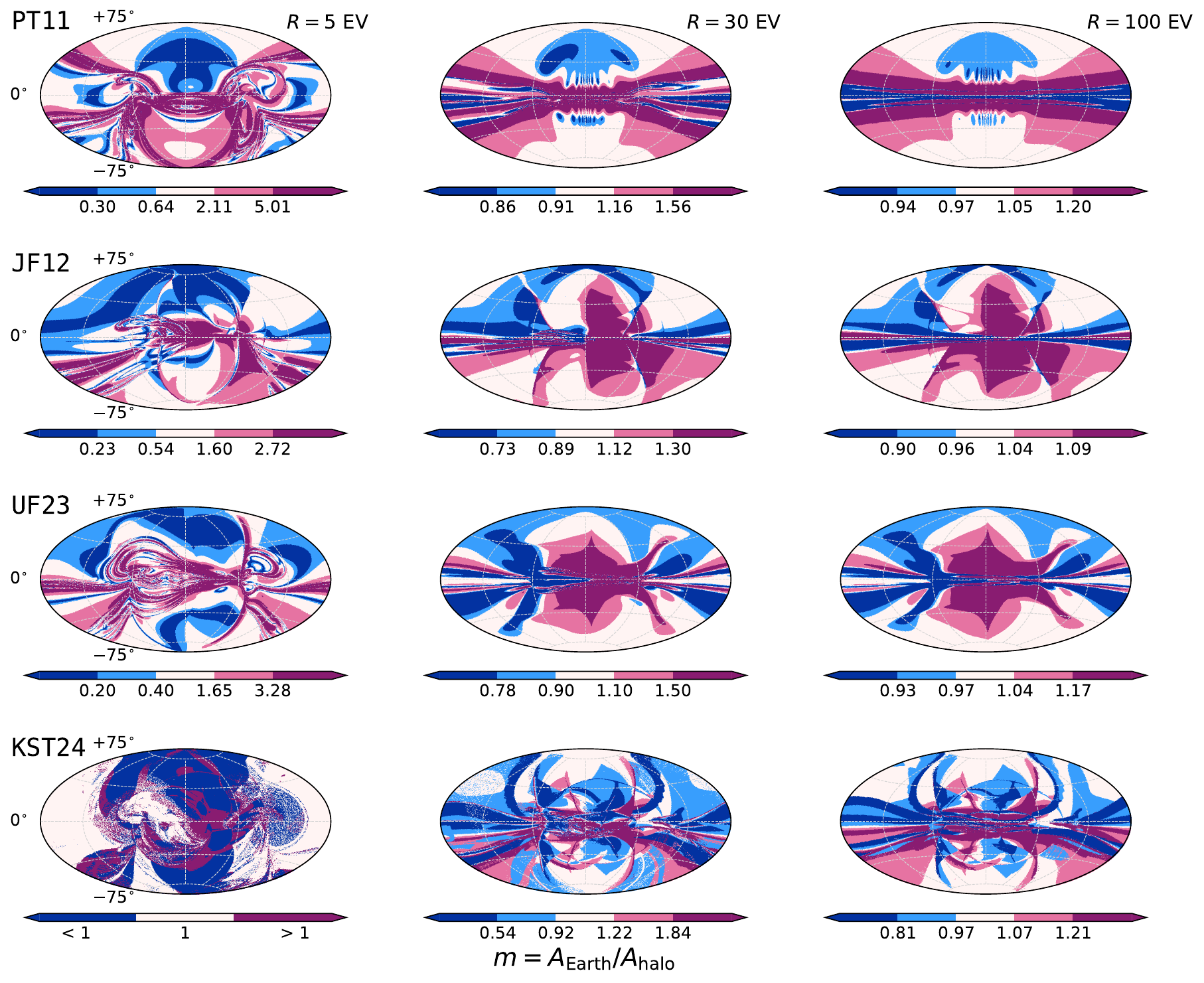}
    \caption{Modification of the CR flux from a source as a function of the arrival direction at Earth for different GMF models and rigidities, shown in Galactic coordinates. Each row displays the results for a distinct GMF model, while each column presents a rigidity value. Note the different ranges for the color scale in each panel.}
    \label{fig:mag}
\end{figure}

For the analysis presented here, we consider 196,886 fiducial particle directions.  Fig.~\ref{fig:mag}, illustrates the results obtained by propagating these fiducial particles and their displacements through the four considered GMF models, at rigidities $R = 5$, 30, and 100\,EV. We note that in the work of Giacinti et al.~\cite{giacinti2010galactic}, the skymaps related to the amplification factor are shown in the halo sky, i.e., outside the Galaxy, while in Fig.~\ref{fig:mag} we show them as a function of the arrival direction at Earth. In Fig.~\ref{fig:mag}, we initially categorized the directions based on whether their ratio was above or below 1. Subsequently, we further divided these into three bins of increasing and decreasing values, ensuring an approximately equal number of directions within each bin. An exception was the lowest rigidity panel for the \texttt{KST24} model, where we used only three bins. This adjustment was made to improve the visualization of the results. It is worth noting that the extreme bins may contain regions with extremely high or low values.

\section{Conclusions}

This work investigated the impact of different Galactic Magnetic Field (GMF) models on the propagation of UHECRs. We considered four GMF models: \texttt{PT11}, the large-scale regular component of \texttt{JF12}, the \texttt{base} variant of \texttt{UF23}, and \texttt{KST24} models. The magnetic deflections imposed by these models can induce phenomena such as the formation of multiple source images and the amplification (or attenuation) of the CR flux from these sources.

We investigated how the magnitude of their deflections changes when propagating the UHECRs through different GMF models. We also explored the appearance of regions with multiple image formation from single extragalactic sources. In particular, we examined the formation of multiple images for two distinct fictitious sources at rigidity $R = 5$\,EV, considering the \texttt{JF12} and the \texttt{UF23} models. Our results demonstrate that the number and directions in the sky of the produced images are highly dependent on the particular GMF model employed. 

Furthermore, we analyzed the modification of the CR flux from a source for different UHECR rigidity values. This was achieved by tracking the evolution of a small area, defined by a fiducial particle and two small orthogonal displacement vectors, as it propagated through the GMF. 

A comprehensive understanding of the GMF is fundamental for the UHECR field, as it profoundly affects the propagation and influences the observables detected on Earth. In the future, we plan to extend this analysis to include the striated and small-scale random fields within the \texttt{JF12} model, as well as incorporating other variants of the \texttt{UF23} model.

\vspace{0.5cm}

\noindent
\large\textbf{Acknowledgments}

\noindent
The authors thank S. Rossoni for useful discussions and comments that improved this work. We acknowledge support by the Bundesministerium für Bildung und Forschung, under grant 05A23GU3.

\newcommand{\etal}{et~al.}
\newcommand{\journal}[5]{\href{https://doi.org/#5}{\textit{#1}\ \textbf{#2} (#3)\ #4}}
\newcommand{\arXiv}[1]{\href{https://arxiv.org/abs/#1}{\nolinkurl{#1}}}


\begin{thebibliography}{99}

\footnotesize\raggedright\setlength{\itemsep}{0pt}

\bibitem{batista2021intergalactic} R. Alves Batista and A. Saveliev, \journal{Universe}{7}{2021}{223}{10.3390/universe7070223}

\bibitem{beck2015spiral} R. Beck, \journal{Astron. Astrophys. Rev.}{24}{2015}{4}{10.1007/s00159-015-0084-4}

\bibitem{auger2015observatory} A. Aab~\etal \ [Pierre Auger coll.], \journal{NIM-A}{798}{2015}{172}{10.1016/j.nima.2015.06.058}

\bibitem{abuzayyad2012telescopearray} T. Abu-Zayyad~\etal \ [Telescope Array coll.], \journal{NIM-A}{689}{2012}{87-97}{10.1016/j.nima.2012.05.079}

\bibitem{aab2017anisotropy} A. Aab~\etal \ [Pierre Auger coll.], \journal{Science}{357}{2017}{1266-1270}{10.1126/science.aan4338}

\bibitem{bakalova2023dipole} A. Bakalová, J. Vícha and P. Trávníček, \journal{JCAP}{2023}{2023}{016}{10.1088/1475-7516/2023/12/016}

\bibitem{bister2024anisotropy} T. Bister, G. R. Farrar and M. Unger, \journal{ApJL}{975}{2024}{L21}{10.3847/2041-8213/ad856f}

\bibitem{hackstein2018anisotropy} S. Hackstein~\etal, \journal{MNRAS}{475}{2018}{2519-2529}{10.1093/mnras/stx3354}

\bibitem{rossoni2025anisotropy} S. Rossoni and G. Sigl \journal{arXiv preprint}{arXiv:2502.19324v2  [astro-ph.HE]}{2025}{}{10.48550/arXiv.2502.19324}


\bibitem{mollerach2025magnetic} S. Mollerach, \journal{PoS}{(UHECR2024)}{2025}{002}{10.22323/1.484.0002}

\bibitem{Abdul2025masscomposition} A. Abdul Halim~\etal \ [Pierre Auger coll.] , \journal{Phys. Rev. Lett.}{134}{2025}{021001}{10.1103/PhysRevLett.134.021001}

\bibitem{castellina2019prime} A. Castellina [Pierre Auger coll.], \journal{EPJ Web Conf.}{210}{2019}{06002}{10.1051/epjconf/201921006002}

\bibitem{batista2022crpropa} R. Alves Batista~\etal, \journal{JCAP}{2022}{2022}{035}{10.1088/1475-7516/2022/09/035}
 
\bibitem{pshirkov2011magnetic}  M. S. Pshirkov~\etal, \journal{ApJ}{738}{2011}{192}{10.1088/0004-637X/738/2/192}

\bibitem{sun2008radio} X. H. Sun~\etal, \journal{A\&A}{477}{2008}{573}{10.1051/0004-6361:20078671}


\bibitem{jansson2012magnetic} R. Jansson and G. R. Farrar, \journal{ApJ}{757}{2012}{14}{10.1088/0004-637X/757/1/14} 

\bibitem{unger2024magnetic} M. Unger and G. R. Farrar, \journal{ApJ}{970}{2024}{95}{10.3847/1538-4357/ad4a54}

\bibitem{korochkin2024magnetic} A. Korochkin, D. Semikoz and P. Tinyakov, \journal{A\&A}{693}{2025}{A284}{10.1051/0004-6361/202451440}

\bibitem{harari1999spectrum} D. Harari, S. Mollerach and E. Roulet, \journal{JHEP}{1999}{1999}{022}{10.1088/1126-6708/1999/08/022}

\bibitem{harari2002lensing} D. Harari~\etal, \journal{JHEP}{2002}{2002}{045}{10.1088/1126-6708/2002/03/045}

\bibitem{farrar2019deflections} G. R. Farrar and M. S. Sutherland, \journal{JCAP}{2019}{2019}{004}{10.1088/1475-7516/2019/05/004}

\bibitem{korochkin2025deflections} A. Korochkin, D. Semikoz and P. Tinyakov, \journal{arXiv preprint}{arXiv:2501.16158v1  [astro-ph.HE]}{2025}{}{10.48550/arXiv.2501.16158}

\bibitem{unger2025deflections} M. Unger and G. R. Farrar, \journal{PoS}{(UHECR2024)}{2025}{003}{10.22323/1.484.0003}

\bibitem{zonca2019healpy} A. Zonca~\etal, \journal{JOSS}{4}{2019}{1298}{10.21105/joss.01298}

\bibitem{gorski2005healpix} K. M. Górski~\etal, \journal{ApJ}{622}{2005}{759}{10.1086/427976}

\bibitem{giacinti2010galactic} G. Giacinti~\etal, \journal{JCAP}{2010}{2010}{036}{10.1088/1475-7516/2010/08/036}

\end{thebibliography}
\end{document}